\documentclass[
   pre,
   superscriptaddress,
   amssymb,
   amsmath,
   superscriptaddress,
   eqsecnum,
   showpacs,
   preprintnumbers,
   byrevtex]{revtex4-1}[10pt]
\usepackage{color}
\usepackage{graphicx}
\begin{document}

\newcommand{\vphi}{\varphi}
\newcommand{\bq}{\begin{equation}}
\newcommand{\be}{\begin{equation}}
\newcommand{\ba}{\begin{eqnarray}}
\newcommand{\eq}{\end{equation}}
\newcommand{\ee}{\end{equation}}
\newcommand{\ea}{\end{eqnarray}}
\newcommand{\tchi} {{\tilde \chi}}
\newcommand{\tA} {{\tilde A}}
\newcommand{\tq} {{\tilde q}}
\newcommand{\tphi} {{\tilde \phi}}
\newcommand{\tp} {{\tilde p}}
\newcommand{\sech} { {\rm sech}}
\newcommand{\pstar}{\mbox{$\psi^{\ast}$}}
\newcommand{\sn} {\text{sn}}
\newcommand{\cn} {\text{cn}}
\newcommand{\dn} {\text{dn}}
\newcommand{\rmd}{\text {d}}                     
\newcommand{\rme}{\text {e}}                     
\newcommand{\Schrodinger}{Schr{\"o}dinger}
\newcommand{\bpsi} {{\bra{\psi}}}
\newcommand{\bphi} {{\bra{\phi}}}
\newcommand{\kpsi} {{\ket{\psi}}}
\newcommand{\kphi} {{\ket{\phi}}}
\newcommand{\calF}{\mathcal{F}}  
\newcommand{\calS}{\mathcal{S}} 
\providecommand{\abs}[1]{\lvert#1\rvert} 
\newcommand{\ket}[1]{\left|{#1}\right\rangle}
\newcommand{\bra}[1]{\left\langle{#1}\right|}
\newcommand{\braket}[2]{\left\langle{#1}\middle|{#2}\right\rangle}
\newcommand{\Braket}[2]{\left\langle{#1}\middle|{#2}\right\rangle}

\preprint{nonlinearQM.tex \today}


\newpage

\title{A Solvable Model of  a Nonlinear extension of Quantum Mechanics}

\author{Alan Chodos}
\email{alan.chodos@uta.edu}
\affiliation{Dept. of Physics, University of Texas at Arlington, 502 Yates Street, Box 19059, Arlington, TX 76019 }

\author{Fred Cooper} 
\email{cooper@santafe.edu}
\affiliation{The Santa Fe Institute, 1399 Hyde Park Road, Santa Fe, NM 87501, USA}
\affiliation{Theoretical Division and Center for Nonlinear Studies,
   Los Alamos National Laboratory,
   Los Alamos, NM 87545}
\date{\today}
\begin{abstract}
We introduce a particular nonlinear generalization of quantum mechanics which has the property that it is exactly solvable in terms of the eigenvalues and eigenfunctions of the Hamiltonian of the usual linear quantum mechanics problem.  We hope that this simple example will elucidate some of the issues of interpreting
nonlinear generalization of quantum mechanics that have been put forth to resolve questions about quantum measurement theory. 
\end{abstract} 
\maketitle

\section{Introduction}

There are a number of reasons to entertain modifications of quantum mechanics. One is simply to push the envelope \cite{Weinberg89,Kaplan22}: is it possible that the \Schrodinger\ equation is an approximation, albeit a very good one, to something else? What could that something else be? 

Another is to try to cure some of the perceived difficulties with quantum mechanics as we know it: can we introduce collapse of the wave function as part of the dynamics without appeal to a separate postulate \cite{Ghiradi86,Pearle89,Bassi13}? Can we avoid probabilities by finding hidden variables \cite{Hance22} that secretly determine the results of measurement, 
without running afoul of Bell's theorem and related restrictions? (Bell said \cite{Bell} ``if [a hidden-variable theory] is local it will not agree with quantum mechanics, and if it agrees with quantum mechanics it will not be local.")     Yet another motivation is the problem of uniting quantum mechanics with a theory of gravity. Most such attempts assume that quantum ideas remain intact and that gravity should be modified. But a minority view \cite{Diosi97,Penrose96,Penrose14} holds that quantum mechanics should also be subject to change. In particular it has been suggested that, to incorporate gravity, quantum mechanics might have to become non-linear. 
To explore some of these issues in a manageable context, in this paper we introduce a particularly simple non-linearity into the \Schrodinger\ equation, which has the advantage that we can completely solve for its effects, whatever the Hamiltonian. However, the non-linearity raises questions of interpretation, which we discuss in the conclusions. Because the interpretation is not settled, we cannot definitively yet say whether our modification suffers from some of the problems of previous attempts to introduce non-linearities, notably, for example, Weinberg's  scheme \cite{Weinberg89} , which was shown, by Gisin \cite{Gisin90} and Polchinski \cite{Polchinski 91}, to involve superluminal communication and other alleged pathologies. 

To proceed, we replace the single \Schrodinger\ equation

\bq \label{se} 
i \frac{\partial}{\partial t} \ket{\psi} = H \ket{\psi}. 
\eq
with a pair of nonlinear  \Schrodinger\ equations
\ba \label{nlse} 
i \frac{\partial}{\partial t} \ket{\psi} &&= H \ket{\psi} + g \ket{\phi} \braket{\phi} {\psi}  \nonumber \\
i \frac{\partial}{\partial t} \ket{\phi} &&= H \ket{\phi} + g^\ast \braket{\psi} {\psi} \ket{\phi}
\ea

Here $g$ is a coupling constant, which we take, in general, to be complex: $g = a + i  b$. Most of our analysis considers the case $b \neq 0$. The $b=0$ case is discussed briefly in the Appendix.

We stress that our equations are valid for any dynamical system to which the Hamiltonian applies, be it non-relativistic quantum mechanics, or relativistic field theory, or something else. As we shall see, assuming the eigenvalue problem for $H$ is solved, we are able to provide complete solutions for $\ket{\psi}$and $\ket{\phi}$.  However, we have assigned a{  \em pair} of state vectors to our system, not just one.  We defer for now the question of how to make use of this embarrassment of riches. 

To relate this nonlinear \Schrodinger\  equation to the nonlinear \Schrodinger\ equations used in the discussion of solitary waves as well as in the 
Gross-Pitaevskii equation (GPE) \cite{Pitaevskii} \cite{Gross}  which describes the behavior of Bose-Einstein Condensates in a trap, it is useful to look at the
one dimensional problem in the ``x'" representation. If we take $H$ to be the free Hamiltonian (for the GPE we would also add an external potential V(x) ) and introduce the notation 
$\psi(x,t)= \Braket { x}{ \psi(t)} $ and. $\phi(x,t)= \Braket {  x}{ \phi(t) } $  ,  we find the equations for $\psi(x,t)$ and $\phi(x,t)$ in one dimension are given by:
\bq
i \frac{\partial \psi(x,t)}{\partial t} =-\frac{1}{2}  \partial_x^2 \psi(x,t)+g \left[\int dx' \phi^*(x',t) \psi (x',t) \right] \phi (x,t),
\eq
and
\bq
i \frac{\partial \phi(x,t)}{\partial t} =-\frac{1}{2}  \partial_x^2 \phi(x,t)+g^* \left[\int dx' \psi^*(x',t) \phi (x',t) \right] \psi (x,t),
\eq
These equations which have a complex but spatially independent ``self-interaction" potential are to be contrasted with the usual NLSE in one dimension (here we add a possible trapping potential as is used in BEC theory) :
\bq \label{NLSE} 
i \frac{\partial \psi(x,t)}{\partial t} = \left( -\frac{1}{2}  \partial_x^2 + V(x) \right) \psi(x,t)+ \lambda |\psi(x,t)|^2 \psi (x,t).
\eq
The latter equation has soliton solutions (as well as Bose Einstein Condensate solutions when we use an appropriate trapping potential).  Several authors have considered using the usual NLSE as a way of introducing how the environment  (external potentials or interactions) modifies our 
usual ideas of having a linear \Schrodinger\ equation and then developing that theme as an {\em alternative} to the usual quantum mechanics.  An example of this is found in \cite{Pang}.  In that particular study, it it suggested that because of interactions with other particles and background potentials in nature, the correct description of a microscopic particle should be the NLSE and {\em not} the linear \Schrodinger\ equation.  The  localized solutions of the NLSE  are then interpreted as the particles of nature which then obey wave-particle duality. 

\section{Solution of the Nonlinear equations} 

 Eqs. (\ref{nlse} ) 
 have one immediate set of solutions: choose two orthogonal vectors,  $\ket{A(t)} $  and $\ket{B(t)}$, each of which solves the original \Schrodinger\ equation. Then  $\ket{\psi} =\ket{A}$  and $\ket{\phi} =\ket{B}$ clearly solve the new equations. This is an exceptional case; we are interested in the general case in which $\Braket{\phi}{ \psi} \neq 0$.

To solve these equations when $\Braket{\phi}{ \psi} \neq 0$. we consider the inner products:
\bq
N = \Braket{\psi}{ \psi} + \Braket{\phi}{ \phi} ; ~\tau = \Braket{\psi}{ \psi} - \Braket{\phi}{ \phi}; ~\gamma = \Braket{\phi}{ \psi}; \delta= \gamma^\ast \gamma. 
\eq
As a consequence of Eqs (\ref{nlse}) and their complex conjugates, these quantities obey equations that are independent of $H$:
\bq \label{timeeqs} 
\frac{dN}{dt}  = 0 ;~  \dot \tau = 4 b \delta ; ~ \dot \gamma  = i ~g \gamma \tau  ;  ~ \dot \delta = -2 b \tau \delta.
\eq

Differentiating the equation for   $\dot \tau$ , and  then using the  equation for $ \dot \delta$, we find that $\tau$ obeys the second order differential equation:
\bq
\frac{d}{d \tau} ( \dot{\tau} + b \tau^2) = 0.
\eq
Integrating once we obtain
\bq
\dot{\tau} + b \tau^2 = b \lambda^2,
\eq
where we have written the constant of integration, which is real as a result of both $\tau$ and $\delta$ being real (as well as $b$), in a suggestive form.  Now for the constant to be real $\lambda$ has to be pure real or pure imaginary.  
We can rewrite the solution to this differential equation formally as:
\bq \label{eqtau}
\int \frac{d \tau}{\lambda^2-\tau^2}  = b  \int dt 
\eq
When $\lambda$ is purely imaginary, $\lambda^2=- \abs {\lambda}^2$  one has that the integral on the left hand side  can be  
written 
\bq
 \int \frac{d \tau}{ \abs {\lambda}^2+ \tau^2}  = -b  \int dt  
 \eq
 Performing the integral using G-R 2.124. \cite{GR80} one obtains.
\bq \frac{1} {\abs{\lambda} } \tan ^{-1}( \frac{\tau}{\abs{\lambda} }) = - b (t -c_2) 
\eq
Inverting one gets the unphysical result (since we want $\tau$ to remain finite)  that:
\bq
\tau =   - \abs {\lambda}  \tan(  b \abs {\lambda}   (t- c_2) ) 
\eq 
Next we will solve the equation in general (but will then just consider real vales of $\lambda$). 
We can rearrange  Eq. (\ref{eqtau}) as

\bq
\int d \tau \left[\frac{1} {\lambda-\tau} +\frac{1} { \lambda+\tau})\right]=(2\lambda b) \int dt.
\eq
Performing the integration one obtains: 
\bq
(\tau+\lambda)/(\tau-\lambda)=e^{2 \xi} 
\eq
where $\xi=\lambda b (t-t'_0) $. Here $ t'_0$ is another possibly complex arbitrary constant.

Solving for $\tau$ algebraically, and then evaluating $ \delta = \dot{ \tau} / (4 b) ,$  we obtain the general solution

\bq
\tau =\lambda  \left[ \frac {e^\xi+e^{-\xi}}{e^\xi-e^{-\xi}}   \right],
\eq
and 
\bq
\delta=-\lambda^2  \frac{1}{ \left[ {e^\xi-e^{-\xi} } \right]^2 } 
\eq
Before we impose the conditions that $\tau$ is real and $\delta$ is  real and positive , this solution depends on 2 arbitrary complex constants, $\lambda $ and $ t'_0.$
The reality condition tells us that $\lambda$ 
must be real or pure imaginary. Furthermore, we can have $t'_0=t_0$ where $ t_0 $is real, or else we can have $t'_0=t_0+i \frac{\pi}{2 \lambda }$ , which will also lead to real values for $\tau$ and $\delta$.  We have already ruled out the case
where $\lambda$ is pure imaginary, since it leads to unphysical blowup of the wave function at finite time.  When $\lambda$ is 
real, the two possible solutions for $\tau$, $\delta$  are with $\lambda = 2 \omega_0 $ 
\ba
\tau &&=2 \omega_0 \tanh (2 b  \omega_0 t) ;~~ \delta=\omega_0^2 ~\text{sech}^2(2 b  \omega_0 t) \nonumber  \\
\tau &&=2 \omega_0 \coth (2 b  \omega_0 t) ;~~ \delta=-\omega_0^2 ~\text{cosech}^2(2 b  \omega_0 t) 
\ea
Here $\omega_0$ is constant, and we have ignored the real constant $t_0$ which just specifies the arbitrary origin of time. 

$\delta$ must be positive, and we want $\tau$ to be non-singular; hence we are limited to the first solution. 
Thus we will assume that our solution is given by 
\bq
\tau =2 \omega_0 \tanh (2 b  \omega_0 t) ;~~ \delta=\omega_0^2 ~\text{sech}^2(2 b  \omega_0 t) 
\eq

Note that we have the relationship  
\bq  \label{omega1} 
\omega_0^2 = \frac{\tau^2}{4} + \delta,
\eq
Below,  for simplicity we will let $\xi= 2 b  \omega_0 t$.
We have that 
\bq
\frac{\dot{ \gamma}}{\gamma} =  i  g \tau = i  2 g   \omega_0 \tanh (2 b  \omega_0 t).
\eq
From the equation for $\gamma$ and using the fact that $\delta= \gamma^*\gamma$ we find that 
\bq \gamma = \omega_0  e^{i \theta} ( \cosh \xi ) ^{i \frac {g}{b} }
\eq

where $\theta$ is an arbitrary phase.  Writing $g=a + i b$ we can  rewrite this as 

\bq \gamma = \omega_0  e^{i \theta}  \sech \xi ( \cosh \xi ) ^{i \frac {a}{b} }
\eq
Note that this satisfies the relationship $\delta= \gamma^*\gamma$. 
If we separate the real and imaginary parts of $\gamma$ we obtain: 
\ba \label{gamma}
\gamma&&= \omega_0  \sech \xi \left(\cos \left( \theta+ \frac{ a \log \left(\cosh ^2 \xi \right)}{2
   b}\right)+i \sin \left( \theta + \frac{a  \log (\cosh ^2 \xi )}{2
   b}\right)\right) \nonumber \\
 &&=   \omega_0 \sech \xi   e^{ i \varphi} = \gamma_1+i \gamma_2,
\ea
where 
$\varphi = \left(\theta + \frac{a \log (\cosh ^2 \xi )}{2 b} \right) $.

It is convenient to parameterize $\ket{\psi}$  and $\ket{\phi}$ as follows:
\ba 
\ket{\psi}&& =  \gamma^{1/2} \sum_n (\hat{\psi}_n(t) e^{- i E_n t} )\ket{n} \\
\ket{\phi}&& =  (\gamma^\ast)^{1/2} \sum_n (\hat{\phi}_n(t) e^{- i E_n t} )\ket{n}
\ea
Here $\ket{n}$ is a complete set of eigenvectors of $H,$ with eigenvalues $E_n$. For notational convenience, we assume the spectrum of $H $is discrete, but this is not necessary.

We find that $\hat{\psi}$ and $\hat{\phi}$ obey the coupled equations

\ba \label{firstorder}
i \dot {\hat{\psi}}_n - \frac{g}{2} \tau  \hat{\psi}_n && = g \hat{\phi}_n |\gamma| = g \hat{\phi}_n \delta^\frac{1}{2}, \nonumber \\
i \dot {\hat{\phi}} _n + \frac{g^\ast}{2} \tau  \hat{\phi}_n && = g^\ast \hat{\psi}_n |\gamma| =g^\ast \hat{\psi}_n \delta^\frac{1}{2}, 
\ea
which after some algebra, lead to the uncoupled second-order equations:
\ba
&&\ddot {\hat{\psi}}_n + g^2 \omega_0^2 \hat{\psi}_n=0  \nonumber \\
&&\ddot {\hat{\phi}}_n + g^{\star 2}  \omega_0^2 \hat{\phi}_n=0.
\ea
The general solution to these equations is 
\ba
\hat{\psi}_n &&= A_n  e^{-\omega_0 b t} e^{i a \omega_0 t }+ B_n e^{+\omega_0 b t} e^{-i a \omega_0 t}  \nonumber \\
\hat{\phi}_n &&= C_n  e^{+\omega_0 b t}e^{i a \omega_0 t} +  D_n e^{-\omega_0 b t} e^{-i a \omega_0 t}.
\ea
We must satisfy the first order equations Eq. (\ref{firstorder})  for $\hat{\phi}_n$ and $\hat{\psi}_n$ and make sure that  $\Braket{\phi}{ \psi} = \gamma$
This determines  $C_n, D_n$.  
\bq
C_n= -  A_n;~~ D_n=  B_n,
\eq

and also imposes the normalization condition:
\bq
\sum_n ( B_n^\ast B_n - A_n^\ast A_n) =1.
\eq
and the orthogonality condition
\bq
\sum_n  B_n^\ast A_n =0.
\eq

It is useful to rescale $A_n,~ B_n$ by $\sinh \vartheta, \cosh \vartheta $ respectively  and then impose the normalizations.
\bq
\sum_n ({A}_n) ^\star A_n= \sum_n ({B}_n) ^\star B_n=1.
\eq
One has that $\vartheta$ is related to the norm $N$:
\bq
N= 2 \omega_0 \cosh[2 \vartheta].
\eq
We can summarize the above as follows.  Let $\ket{A}$ and $\ket{B}$ be two normalized orthogonal solutions of the original  \Schrodinger\ equation.
i.e.
\bq
\ket{A} = \sum_n {A}_n e^{-i E_n t} \ket{n} ;~~\ket{B} = \sum_n {B}_n e^{-i E_n t}  \ket{n} 
\eq
Then
\ba \label{explicit} 
\ket{\psi} &&= \gamma^{1/2} \left[ e^{i g \omega_0 t}  \sinh \vartheta \ket{A} +  e^{-i g \omega_0 t}   \cosh \vartheta \ket{B} \right] \nonumber \\
\ket{\phi} &&= {\gamma^\ast}^{1/2} \left[ -e^{i g^\ast \omega_0 t}  \sinh \vartheta \ket{A} +  e^{-i g^\star \omega_0 t}   \cosh \vartheta \ket{B} \right]  
\ea
Note that  $\gamma= \Braket{\phi}{ \psi}$ goes to zero as $ t \rightarrow \pm \infty$, so that in those regions we expect that $\ket{\psi}$ and $\ket{\phi}$ obey 
the usual  \Schrodinger\ equation.  In fact we find that indeed as $t \rightarrow - \infty$, 
\ba
\ket{\psi} &&\rightarrow \gamma^{1/2} e^{i g \omega_0 t}  \sinh \vartheta \ket{A}  \rightarrow \sqrt{2 \omega_0} e^{i \hat{\theta}/2 }  \sinh \vartheta \ket{A}   \nonumber \\
\ket{\phi} &&\rightarrow  {\gamma^\ast}^{1/2}  e^{-i g^\star \omega_0 t}   \cosh \vartheta \ket{B}  \rightarrow \sqrt{2 \omega_0} e^{-i \hat{\theta}/2}  \cosh \vartheta \ket{B}   . 
\ea

whereas as $t \rightarrow + \infty$, 
\ba
\ket{\psi} &&\rightarrow \gamma^{1/2} e^{-i g \omega_0 t}  \cosh \vartheta \ket{B}  \rightarrow \sqrt{2 \omega_0} e^{i \hat{\theta}/2}  \cosh \vartheta \ket{B}   \nonumber \\
\ket{\phi} &&\rightarrow  -{\gamma^\ast}^{1/2}  e^{-i g^\star \omega_0 t}   \sinh \vartheta \ket{A}  \rightarrow -\sqrt{2 \omega_0} e^{-i  \hat{\theta}/2 }  \sinh \vartheta \ket{A}   . 
\ea
We see we can write this in terms of an $S$ matrix  which is unitary and which takes the state at $t \rightarrow - \infty$ to the state at $ t \rightarrow +\infty$.
We find
\bq
\left(
\begin{array}{c}
 \ket{\psi}_+ \\
 \ket{\phi}_+  \\
\end{array}
\right) = 
S  
\left(
\begin{array}{c}
 \ket{\psi}_- \\
 \ket{\phi}_-  \\
\end{array}
\right)  
\eq 
where 
\bq
S =  \left(
\begin{array}{cc}
 0 & e^{i \hat{\theta}} \\
- e^{-i \hat{\theta} } & 0 \\
\end{array}
\right)
\eq
where $\hat{\theta} = \theta- \frac{a}{b} \ln 2$.
We can now introduce the density matrix $\rho$ via

\bq \label{rho} 
\rho =\frac{1}{N} \left( \kpsi \bpsi+\kphi \bphi \right)
\eq 
which satisfies $Tr \rho =1$.
 Unlike the individual $\ket{\psi}$ and $\ket{\phi}$, $\rho$ satisfies the linear evolution equation of the usual \Schrodinger\ equation.
 \bq
 i \frac{\partial \rho}{\partial t} = [H, \rho].
 \eq
 Using the explicit forms of $\ket{\psi}$ and $\ket{\phi}$ Eqs. (\ref{explicit})
 we obtain 
 \bq
 \rho = \frac{2 \omega_0}{N} \left[ \sinh^2 \vartheta \ket {A} \bra{A} + \cosh^2 \vartheta \ket{B} \bra{B} \right].
 \eq
 Thus we see that the eigenstates of $\rho$ are the vectors $\left(\ket{A} ,\ket{B} \right)$ with eigenvalues $\frac{ 2\omega_0}{N} \left(   \sinh^2 \vartheta,\cosh^2 \vartheta \right)$. 
 
We find by straightforward calculation that
\bq
Tr \rho^2 = 1- \frac{2 \cal S }{N^2}
\eq
where the positive  quantity $\cal S$, called the Schwarz parameter is given by: 
\ba
\cal S &&= \braket {\psi}  {\psi}  \braket {\phi} {\phi} - \braket{\psi} {\phi} \braket{\phi}{\psi}  \nonumber \\
&&= 4 \omega^2_0 \left(\sinh^2 \vartheta \cosh^2 \vartheta \right).
\ea
which is clearly independent of time.  We see that $\cal S$ can vanish if and only if $\ket{\psi}$ and $\ket{\phi}$ are proportional which requires  $\vartheta=0$. In that case both $\ket{\psi}$ and $\ket{\phi}$ are proportional to $\ket B$ and the system is in a ``pure"  state.  Because $\cal S$ is a constant, $\ket{\psi}$ and $\ket{\phi}$ are always proportional or they never are.

\section{Lagrangian Formalism}
One of the nice properties of the usual NLSE is that it can be derived from Dirac's Action Principle \cite{Dirac} , which leads to 
Lagrange's equations for the fields. That is, 
an action that upon variation leads to the equation of motion Eq. (\ref{NLSE}) is given by (see for example \cite{CLS}) 
\begin{subequations}\label{DiracAction}
\begin{align}
   \Gamma[\psi,\psi^{\ast}]
   &= \int dt ~ L[\psi,\psi^{\ast}]
   \label{DiracAction-a} \\
   L[\psi,\psi^{\ast}]
   &=
   \frac{i}{2} \int dx~
   \bigl [\,
      \psi^\ast (\partial_t \psi) - (\partial_t \psi^\ast) \psi \,
   \bigr ]
   - 
   H[\psi,\psi^{\ast}] \>,
   \label{DiracAction-b} \\
   H[\psi,\psi^{\ast}]
   &=
 \int dx ~
   \Bigl [\,
      \frac{1}{2} \, | \psi_x |^2
      + 
      \frac{\lambda }{2} \, (\psi^\ast \psi)^2
      +
      V \, | \psi |^2 \, 
   \Bigr ] \>,
   \label{DiracAction-c} 
   \end{align}
\end{subequations}
The equations of motion arise from the variations:
\bq
\frac{\delta \Gamma}{\delta \psi}  = \frac{\delta \Gamma}{\delta \psi^\star} =0. 
\eq

We have not been able to find a Lagrangian that  will yield eqs.  (\ref{nlse})  directly, and we are not sure if one exists. 
However, once we have solved for $\gamma(t)$ we can linearize eqs. (\ref{nlse}) to obtain 

\bq \label{nlselin} 
i \frac{\partial}{\partial t} \ket{\psi} = H \ket{\psi} + g \gamma  \ket{\phi} 
\eq
\bq
i \frac{\partial}{\partial t} \ket{\phi} = H \ket{\phi} + g^\ast  \gamma^\ast  \ket{\psi} 
\eq
This  version of our equations  can be derived from the action:

\ba  \label{action1} 
\Gamma &&= \int dt~  \cal{L}  \nonumber \\
&&=  \int dt~  
 \left[  \bra{\psi}\left( i \frac{\partial}{\partial t} -H \right)  \ket{\psi} 
 +   \bra{\phi}\left( i \frac{\partial}{\partial t} -H \right)  \ket{\phi}  \right. \nonumber \\
&&\left.  - (r_1+i r_2)  \braket{\psi}{\phi} - (r_1- i r_2)   \braket{\phi}{\psi} \right] ,
\ea
from the equations: 
\bq \label{var}
 \frac{\delta \Gamma}{\delta \bra{\psi}} =0 ;~~
 \frac{\delta \Gamma}{\delta \bra{\phi}} = 0.
  \eq

Here $g \gamma = r_1+i r_2= (a \gamma_1- b \gamma_2) + i (b \gamma_1+a \gamma_2 ) $, and $\gamma_1, \gamma_2$ are given in 
Eq. (\ref{gamma}).


\section{Dissipation Function Formulation}
Although the linearized equations have a Lagrangian formalism,  if we naively generalize  Eq. (\ref{action1} ) we obtain a real action which only gives
the real part of the full equations.  That is the naive generalization is

\ba 
\Gamma &&= \int dt~  \cal{L}  \nonumber \\
&&=  \int dt~  
 \left[  \bra{\psi}\left( i \frac{\partial}{\partial t} -H \right)  \ket{\psi} \right. \nonumber \\
&& \left. +   \bra{\phi}\left( i \frac{\partial}{\partial t} -H \right)  \ket{\phi} 
 - (g+g^\ast)/2  \braket{\phi}{\psi} \braket{\psi} {\phi} \right] ,
\ea
which only depends on the the real part of $g$. 
 
Since the nonlinear terms in our equation are similar to  a complex self consistent potential,   the equations  we are studying here share some properties with the usual 
nonlinear \Schrodinger\ equation  \cite{Zakharov} \cite{Sulem} in an
external complex potential. That system has been studied in a collective coordinate approximation using a Dissipation Functional Formalism \cite{Mertens} . 
At least formally we can derive our nonlinear complex equations using this formalism. 
Again writing the complex coupling constant here as $ g= a+ i b$,
we find now that the real part of Eqs. (\ref{nlse}) can be obtained from the simple action:
\ba 
\Gamma &&= \int dt~  \cal{L}  \nonumber \\
&&=  \int dt~  
 \left[  \bra{\psi}\left( i \frac{\partial}{\partial t} -H \right)  \ket{\psi} \right. \nonumber \\
&& \left. +   \bra{\phi}\left( i \frac{\partial}{\partial t} -H \right)  \ket{\phi} 
 - a \braket{\phi}{\psi} \braket{\psi} {\phi} \right] ,
\ea
by varying this action with respect to $\bra{\psi}$ and $\bra{\phi} $.
In analogy with \cite{Mertens},  we find that 
there is a dissipation functional
\bq
F = \int dt ~\calF 
\eq
such that  the full  Eqs. (\ref{nlse}) can be obtained from: 
 \ba \label{diss}
 \frac{\delta \Gamma}{\delta \bra{\psi}} &&= - \frac{\delta F}{\delta \bra{\psi_t} }\nonumber \\
 \frac{\delta \Gamma}{\delta \bra{\phi}} &&= - \frac{\delta F}{\delta \bra{\phi_t} }
  \ea
  Here we use the notation $ \bra{\phi_t} = \partial_t \bra{\phi}$ etc. 
In the $``x"$  representation  this would translate to 
\bq
\frac{\delta \Gamma}{\delta \bra{ \psi_t}} \rightarrow  \frac{\delta \Gamma}{\delta \psi_t^\star(x,t) }.
\eq
  The appropriate $\calF$ is given by 
  \bq
\calF = - i  {b }  \left [ \braket{\phi}{\psi}   \braket {\psi_t}{\phi}  - \braket{\phi} {\psi_t } \braket {\psi } { \phi}  -\braket{\psi}{\phi}   \braket {\phi_t}{\psi} +  \braket{\psi} {\phi_t } \braket {\phi } { \psi}  \right]
\eq

The complex conjugate of  Eqs. (\ref{nlse}) can be obtained from the complex conjugate of Eqs. (\ref{diss}.)
So at least formally we can derive  Eqs. (\ref{nlse}) using this formalism.  This formalism is very useful if one wants to study the time evolution of an initial state in a collective coordinate approximation such as a Time-Dependent Hartree approximation and its generalizations.

\section{Discussion} 

In ordinary quantum mechanics, a dynamical system is described by a state vector evolving according to a particular Hamiltonian. In our modified version, in contrast, we need two orthogonal such state vectors, $\ket{A}$ and  $\ket{B}$ , and the evolution is given by a linear combinations of these, $\ket{\psi}$  and $\ket{\phi}$ with time-dependent coefficients that depend on 2 additional parameters, $\omega_0$ and $\vartheta$. So a given dynamical system is somewhat more intricate than ordinary quantum mechanics would lead us to believe.

In order fully to determine either $\ket{\psi}$  or $\ket{\phi}$  we need to specify their asymptotic values both as $t \rightarrow -\infty$ and as $t \rightarrow +\infty$. Once we so specify one of $\ket{\psi}$  or $\ket{\phi}$ ,  the other is then determined. This circumstance has some similarities with the work of Aharonov, Bergmann and Lebowitz (ABL)  \cite{Aharonov64} , and also \cite{Aharonov-Vaidman}, who consider systems that are both pre- and post-selected, in the context of ordinary quantum theory, and ask for the resulting probabilities if an observable is measured at some intermediate time $t$. 

From an operational point of view, though, it seems much more natural to specify the system at some initial time, and then to give a recipe for the outcome of a measurement at some later time, without having to know the state of the system in the far future. But then, in our formulation, we have to specify twice as much information as is normally given in ordinary quantum mechanics. We need to know what the significance of this extra information is. If an experimentalist prepares the system, say in its ground state, what does this mean in terms of $\ket{\psi}$  and $\ket{\phi}$? If both $\ket{\psi}$ and  $\ket{\phi}$ need to be in the ground state, then we are demanding $\vartheta =0$ . But maybe that is too restrictive. The proper rule for mapping our formalism onto a physical system is a major unresolved issue.

It would be of great interest to explore the consequences of our modification of quantum mechanics and to show how these nonlinear effects can be observed in macroscopic systems. However, given the uncertainties in interpretation caused by having two state vectors instead of just one, we are unable to address this issue here. Our coupling constant $g$ has units of energy, so that as long as $ \abs{g}$  is small compared to the typical energy scale of the Hamiltonian, we expect that the non-linear effects will be similarly small. Exactly how to search for these effects is a question that must be relegated to future work.

One possible scenario is that the extra information encoded in  $\ket{\psi}$  and $\ket{\phi}$ can be used to make the result of a measurement deterministic. Then the probabilistic rules of quantum mechanics would emerge when averaging over the extra information. It remains to be seen whether such a scenario can be realized. If it can, that might open the way to coupling this modified form of quantum mechanics to a purely classical form of general relativity, since in a deterministic theory collapse of the wave function is avoided.

On a more poetic note, one could imagine that $\ket{\psi}$  and $\ket{\phi}$ exist in separate universes, so that we live, for example, in the  $\ket{\psi}$ universe, which communicates with the $\ket{\phi}$ universe via the non-linear terms. While our universe evolves from $\ket{A}$ to $\ket{B}$ the $\ket{\phi}$ universe evolves the other way, from $\ket{B}$  to $\ket{A}$ (or evolves the same way, but backwards in time). Related to this, we observe that if there is an anti-unitary operator that commutes with $H$, then our equations (\ref{nlse})  enjoy a time-reversal symmetry, which involves both $t$ going to $ -t$ and the interchange of $\ket{\psi}$ and $\ket{\phi}$.

\appendix
\section {The Case of a Single State Vector}

Even simpler than the case we consider in the main text is when we have only one state vector, coupled non-linearly to itself, that is 
\bq  \label{a1}
i \frac{\partial}{\partial t}  \ket{\psi} =H \ket{\psi} +g \ket{\psi} \Braket{\psi}{ \psi} .                          
\eq
From Eq. (\ref{a1})  and its complex conjugate, we immediately derive
\bq \label{a2} 
i \frac{\partial}{\partial t} \Braket{\psi}{ \psi}        =(g-g^* ) \Braket{\psi}{ \psi}^2 .                      
\eq

If  $g$ is real, then $ \Braket{\psi}{ \psi} =N$  where $N$ is a constant. In that case, the general solution is just
\bq
\ket{\psi}= \sum_n  c_n e^{-i(E_n+gN) t}   \ket{n}.
\eq 

where the $ (\ket{n} )$ are eigenstates of $H$ with eigenvalues  $E_n $, and the $c_n $ are constants obeying 
\bq
\sum_n |c_n |^2=N. 
\eq

If $g$ is complex, $g=a+ib$ then the solution to Eq. (\ref{a2})  is 
\bq
\Braket{\psi}{ \psi}    = \frac {-1} {2b (t-t_0 ) }     .
\eq
This only makes sense if $ 2b (t-t_0 ) < 0 $. If  $b > 0$ , the system ceases to exist after $t=t_0$. Conversely, if $b<0$ , the system comes into existence with infinite norm at $t=t_0$, and the norm decays inversely with $ t $ thereafter. These pathologies are associated with the fact that, unlike for the two state-vector case, the self-coupling is non-hermitian if $g$ is complex. In the case $ b<0$  this behavior is reminiscent of the evolution of the universe from an initial big bang.

As in the two state-vector case, one can go on to construct the general solution for $\ket{\psi}$  in the form of a function of time multiplying a solution of the usual \Schrodinger\  equation. 

\section{ The case  $g$ is real  ($b=0$)}
In this case as we have shown there is a Lagrangian description of the problem.  
In the main text we have concentrated on the general case of complex $g$. It is not straightforward to obtain the solution to the special case of real $g$ by setting $ b=0 $ in the solution. Rather, we return to equations (\ref{timeeqs}) and note that when $b=0$, $\tau$ as well as $\delta$ are  constants:
\bq
\tau = \tau_0= const.
\eq

which leads to an immediate solution for $\gamma = \Braket{\phi}{\psi}$
\bq
\gamma= \gamma_0 e^{i g \tau_0  t} 
\eq
where $\gamma_0$  is a constant.  Note that this leads to $\delta = |\gamma_0|^2$.  As in the general case, it is convenient to parameterize the state vectors as

\ba 
\ket{\psi}&& =  \gamma^{1/2} \sum_n (\hat{\psi}_n(t) e^{- i E_n t} )\ket{n} \\
\ket{\phi}&& =  (\gamma^\ast)^{1/2} \sum_n (\hat{\phi}_n(t) e^{- i E_n t} )\ket{n}
\ea

which imply the equations

\ba \label{firstorder2}
i \dot {\hat{\psi}}_n - \frac{g}{2} \tau_0  \hat{\psi}_n && = g \hat{\phi}_n |\gamma_0| , \\
i \dot {\hat{\phi}} _n + \frac{g}{2} \tau_0  \hat{\phi}_n && = g \hat{\psi}_n |\gamma_0|  
\ea
which after some algebra, lead to the uncoupled second-order equations:
\ba
&&\ddot {\hat{\psi}}_n + g^2 \omega_0^2 \hat{\psi}_n=0  \nonumber \\
&&\ddot {\hat{\phi}}_n + g^{2}  \omega_0^2 \hat{\phi}_n=0,
\ea
where 
\bq  \label{omega} 
\omega_0^2 = \frac{\tau_0^2}{4} + \delta,
\eq

The general solution to these equations is 
\ba
\hat{\psi}_n &&= A_n e^{i g \omega_0 t }+ B_n e^{-i g \omega_0 t} \nonumber \\
\hat{\phi}_n &&= C_n e^{i g \omega_0 t} +  D_n e^{-i g \omega_0 t}
\ea
where
\bq
C_n= - \frac{1}{k} ~~ A_n;~~ D_n=  k B_n,
\eq
and
\bq
 k = \frac{\sqrt{\delta}}  {\omega_0 + \frac{\tau_0}{2} }= \frac{(\omega_0 -\tau_0 /2)  } {\sqrt{\delta}}.
 \eq
The two forms of $k$ are equivalent because of the definition of $\omega_0$. 
The requirement that  $\Braket{\phi}{ \psi} = \gamma$ leads to the conditions
\bq
\sum_n ( k~B_n^\ast B_n - \frac{1}{k} A_n^\ast A_n) =1.
\eq
and the orthogonality condition
\bq
\sum_n  B_n^\ast A_n =\sum_n  A_n^\ast B_n=  0.
\eq

\acknowledgments
We are grateful to Sean Carroll and David Fairlie for conversations. AC would like to thank the Santa Fe Institute for hospitality while part of this work was done.

\end{document}